\documentclass[10pt]{article}
\usepackage{inputenc}
\usepackage{amsmath}
\usepackage{amssymb, amsfonts}
\usepackage{graphicx,epstopdf,bm}
\usepackage[colorlinks,bookmarks,urlcolor=blue,citecolor=blue,linkcolor=blue]{hyperref}
\usepackage[paperwidth=210mm, paperheight=297mm, textwidth=350pt, marginparsep=30pt, marginparwidth=124pt, textheight=692pt, footskip=60pt]{geometry}
\usepackage{listings}
\usepackage[T1]{fontenc}
\DeclareMathOperator*{\arginf}{arg\,inf}
\newcommand{\pd}[2]{\frac{\partial #1}{\partial #2}}

\newcommand{\pcd}[3]{\frac{\partial^{2} #1}{\partial #2 \partial #3}}

\newcommand{\abs}[1]{\left\vert #1 \right\vert}
\newcommand{\ave}[1]{\left \langle #1 \right \rangle}
\newcommand{\explr}[1]{\exp\left\{ #1 \right\}}
\DeclareMathOperator*{\diagop}{diag}
\newcommand{\diag}[1]{\diagop \{ #1 \} }
\newcommand{\chem}[2]{ {{#1\atop\longrightarrow}\atop{\longleftarrow\atop #2}} }

\newcommand{\rmp}{P}
\newcommand{\Dt}{\Delta t}

\newcommand{\act}{\mathcal{S}}
\newcommand{\vx}{\mathbf{x}}
\newcommand{\vp}{\bm{p}}
\newcommand{\vxa}{x_{A}}
\newcommand{\lsim}{\stackrel{\ln}{\simeq}}
\newcommand{\bfone}{\mathbf{1}}

\begin{document}
\title{Asymptotic and numerical methods for metastable events in stochastic gene networks}
\author{Jay Newby\thanks{Mathematical Bioscience Institute, Ohio State University, 1735 Neil Ave. Columbus, OH 43210}}
\maketitle
\begin{abstract}
A general class of stochastic gene expression models with self regulation is considered. One or more genes randomly switch between regulatory states, each having a different mRNA transcription rate. The gene or genes are self regulating when the proteins they produce affect the rate of switching between regulatory states. Under weak noise conditions, the deterministic forces are much stronger than fluctuations from gene switching and protein synthesis. Metastable transitions, such as bistable switching, can occur under weak noise conditions, causing dramatic shifts in the expression of a gene. A general tool used to describe metastability is the quasi stationary analysis (QSA). A large deviation principle is derived so that the QSA can explicitly account for random gene switching without using an adiabatic limit or diffusion approximation, which are unreliable and inaccurate for metastable events.This allows the existing asymptotic and numerical methods that have been developed for continuous Markov processes to be used to analyze the full model.
\end{abstract}

\section{Introduction}


Stochasticity in gene regulation circuits is known to cause rare extreme shifts in the expression of a gene, which can have a profound effect on the behavior of a cell \cite{eldar10a,maheshri07a}.
This leads to the question of how cells coordinate and regulate different sources of biochemical fluctuations, or noise, to function within a genetic circuit.
Rare, noise-induced dynamical shifts in a stochastic process are known as {\em metastable events}.
For example, Brownian motion in a double well potential, where the fluctuations are weak compared to the force of the potential, displays bistable switching.
In general, metastable events occur when fluctuations are weak compared to deterministic forces, and the stochastic process is said to be under {\em weak noise} conditions.
The chalenge for stochastic modeling is to predict and explain the possible metastable behaviors and offer a testable hypothesis by quantifying the timescale on which those events are likely to occur.

Standard methods for computing transition rates in chemical reactions cannot be applied to a gene expression model when it includes reactions that involve one or more genes.
For chemical systems with a large number $N$ of molecules, the stochastic model converges to the mass action model in the large $N$ limit.
Transition rate approximations for chemical systems generally apply to weak noise conditions where $N$ is large but finite.
A deterministic mass action description is not valid for a gene expression model because there can be as few as one copy of a given gene.
Hence, we cannot assume weak noise conditions by assuming that $N$ is large.
Even if we can find a way to define weak noise for the gene expression model, the standard transition rate approximation does not apply.

A deterministic description can be obtained using a separation of time scales if the gene reactions are fast compared to other reactions; this is known as an adiabatic limit.
For example, a gene that rapidly switches between on and off states would, in the adiabatic limit, have an effective transcription rate scaled by the fraction of time spent in the on state.
A stochastic gene regulation model can then be said to be under weak noise conditions when gene switching is fast but not infinitely fast.

Thus, there is a separation of timescales, where gene switching is the fast reaction and protein synthesis/degradation is the slow reaction.
Almost all of the approaches to date resolve the gene switching problem by either taking the adiabatic limit \cite{aurell02a} or using a diffusion approximation \cite{aurell02a,walczak05a,feng14a,lu14a}, but diffusion approximations of metastable phenomena, in general, are known to be highly unreliable, even for relatively simple birth-death processes \cite{doering07a}.
The adiabatic diffusion approximation (also known as stochastic averaging or stochastic quasi-steady state) can be particularly unreliable \cite{newby11b,keener11a,newby13a}.
Other approaches are either not generally applicable \cite{assaf11a} or result in a difficult to solve high dimensional problem \cite{sasai03a,zhang14a}.
We propose an accurate approximation that exploits the separation of timescales without ignoring gene switching or using diffusion approximations of any kind.

The study of metastable behavior in probability theory is known as {\em large deviation theory} \cite{freidlin12a,feng06a}, and has many mathematical similarities to the path integral description of quantum field theory.
Large deviation theory is mathematically rigorous and establishes an important link between the path distribution and the stationary probability density function, but for most chemical systems and continuous Markov processes, the WKB approximation is the simplest and most direct means of calculating certain weak noise approximations.

The Quasistationary analysis (QSA) uses the WKB method to approximate the stationary probability density function, which also yields information about the long time behavior of the process \cite{schuss10a}.
One useful biproduct of the WKB approximation is the so-called {\em most probable path} (MPP).
Under weak noise conditions, the probability that the process takes a particular path from point A to point B is sharply peaked along a MPP.~ 
The likelihood of deviating from the MPP is an exponentially decreasing function of the magnitude of the deviation.
In other words, stochastic trajectories are highly likely to closely follow the MPP.~ 
A MPP is a statistic, similar to the mode, of the probability distribution of stochastic trajectories, (i.e., a probability distribution in a function space).
Although they describe a stochastic process, MPPs themselves are not stochastic.
In general, there are two kinds of MPPs.~ 
If a deterministic trajectory connects point A to point B, then the MPP is the deterministic trajectory. 
If the transition is noise induced, such as a metastable transition, MPPs build {\em action} and become more improbable as the path gets longer.
MPPs can tell us how different metastable transitions occur; they can be thought of as describing noise induced dynamical behavior.
By combining the stationary density and MPP approximations, the mean first exit time for a metastable events (such as bistable switching) can also be approximated using a spectral projection method.
To summarize, the QSA is based on the WKB approximation and can be used to obtain asymptotic approximations of: (i) the stationary probability density function, (ii) MPPs, and (iii) mean first exit times for metastable transitions.
How to apply the QSA to processes that have both large-$N$-type species and fast (adiabatic) species with low copy number is an unsolved problem.

Wolynes and coworkers showed how the quantum mechanical path integral formulation of spin boson systems could be applied to study metastability in stochastic gene expression models that include gene switching \cite{sasai03a,zhang14a}.
They obtained a path integral description in the full space of variables (both slow and fast) \cite{zhang14a}, but the result is a high dimensional problem that can be difficult to solve.
In this paper, we present an approach that exploits the separation of time scales.
Recently, the QSA has been adapted to get approximations of the stationary density and mean switching times \cite{newby12a,newby13a,newby14mlpp}, but the link to a path integral formulation is ambiguous because the WKB approximation does not uniquely determine a Hamiltonian.
(The leading order problem from the WKB expansion is a static Hamilton-Jacobi equation.)
Two important numerical algorithms (the geometric minimum action method \cite{heymann08a,vanden-eijnden08a} and the ordered upwind method \cite{cameron12a,newby14mlpp}) critically depend on the correct formulation of the Hamiltonian.

Recent advances in this area have been made using probability theory.
Using a {\em large deviation principle}, one can derive the action functional, which is the basis for approximations of metastable behavior.
Freidlin and Wentzell \cite{freidlin12a} present a rigorous proof of a large deviation principle for the case where the fast process does not depend on the slow variables (i.e., no gene regulation), and Kifer \cite{kifer09a} extended the result to the fully coupled case without noise in the slow variables (the slow process by itself is deterministic).
The model is called {\em piecewise deterministic} if the fast process is a discrete Markov process and the slow process is deterministic.
A large deviation principle was developed for piecewise deterministic processes using path integrals \cite{bressloff14a}.
The piecewise deterministic case is relevant for a gene expression model in the large $N$ limit where the protein concentration changes deterministically in between random gene state transitions \cite{newby12a}.
In this paper, we consider the case where the slow process also has noise from protein synthesis and degradation.
Specifically, we assume that the synthesis rate is much larger than the degradation rate so that the protein is under large-$N$-type weak noise conditions and can be treated as a continuous concentration.

One of the weaknesses of the large deviation theory approach is that the resulting approximations are logarithmically asymptotic.
For example, if $P_{\epsilon}(x)$ is a probability density function that depends on a small parameter $\epsilon \ll 1$, a large deviation principle yields an approximation of the form, $\epsilon\log(P_{\epsilon}(x)) \sim - \Phi(x) + O(\epsilon)$, which we write as $P_{\epsilon}(x) \lsim e^{-\Phi(x)/\epsilon}$.
The QSA yields an asymptotic approximation that includes a pre exponential factor $K(x)$ so that $P_{\epsilon}(x) \sim K(s)e^{-\Phi(x)/\epsilon}$.

In this paper, we show how to resolve the ambiguity in the WKB approximation, which allows the QSA to be applied to models of stochastic gene networks that include multiple types of weak noise (large-$N$-type weak noise and adiabatic-type weak noise).
For simplicity we ignore the dependence on mRNA copy number and focus on gene state switching and protein synthesis and degradation.
The author has developed the QSA for single gene models that include mRNA using transform methods \cite{newby14mrnapp}, and we anticipate that the theory can be expanded to include gene networks in future work.
We develop a path description that takes advantage of the separation timescales without using an adiabatic reduction (stochastic averaging) or any other form of diffusion approximation.
This can be thought of as an averaging-like procedure that yields highly accurate approximations of many relevant quantities, including the distribution of the fast gene state.
We then combine the results of large deviation theory and the QSA.
That is, for practical application, we show how to augment the WKB method to obtain all relevant quantities without the need to consider path integrals.

The paper is organized as follows. 
First, we describe a general stochastic model of gene expression in Section \ref{sec:model}.
Then, in Section \ref{sec:wkb} we present the WKB approximation and explain how to resolve the ambiguity in defining the Hamilton--Jacobi equation.
In Section \ref{sec:ldp} we show how a single mathematical concept (the asymptotic approximation of an integral using Laplace's method) can be used to derive a large deviation principle, which establishes how to correctly apply the WKB approximation.
Finally, in section \ref{sec:tog} we illustrate the results with the simple example of the {\em toggle switch} model (also known as the {\em mutual repressors} model) \cite{kepler01a,newby12a}.

\section{Model}
\label{sec:model}
Let $s = 1,\cdots, N_{s}$ label the state of one or more genes.
For simplicity, we assume that protein synthesis is directly coupled to the gene state and ignore mRNA (for more on applying the QSA to a model that does consider mRNA see \cite{assaf11a,newby14mrnapp}.)
Assume that the gene or genes are responsible for synthesizing $d>0$ protein species $X_{1}, \cdots, X_{d}$, and let $\vx\in \mathcal{D} \subset \mathbb{R}^{d}$ be the vector of protein concentrations.
Protein species $X_{k}$ is synthesized and degraded according to the reaction,
\begin{equation}
  \emptyset \chem{\sigma_{k}(s)/\epsilon}{\gamma_{k}} X_{k},
\end{equation}
where $\sigma_{k}(s)/\epsilon$ is the gene state dependent production rate and $\gamma_{k}$ is the degradation rate.
Assume that $0 < \epsilon \ll 1$ is a small parameter so that $\vx$ can be treated as a continuous variable.
We assume that there are no direct interactions between the different protein species, but extending the analysis to account for such reactions is straightforward.

Given a fixed gene state $s$, the Master equation describing the evolution of the probability density function $P(\vx, t)$ is
\begin{multline}
  \label{eq:3}
  \pd{}{t}P(\vx, t) = L^{\epsilon}_{\vx}P \equiv \frac{1}{\epsilon}\sum_{k = 1}^{d}\sum_{x_{k}'\in \mathcal{D}}\left[\sigma_{k}(s)(\delta_{x_{k}', x_{k} - \epsilon} - \delta_{x_{k}', x_{k}}) \right.\\ 
\left. + \gamma_{k} (\delta_{x_{k}', x_{k} + \epsilon} - \delta_{x_{k}', x_{k}})x_{k}' \right]P(\vx', t)
\end{multline}
Assume that the $X_{1}, \cdots, X_{d}$ regulate the activity of the gene or genes so that the transition rates are functions of $\vx$.
Given a fixed concentration $\vx$, the gene state is described by the Master equation
\begin{equation}
  \label{eq:5}
  \pd{}{t}P(s, t) = \frac{1}{\epsilon}A^{\vx}P \equiv \frac{1}{\epsilon}\sum_{s'=1}^{N_{s}}A^{\vx}_{ss'}P(s', t),
\end{equation}
where $A^{\vx}$ is the transition rate matrix, assumed to be irreducible so that there is a steady-state distribution $\rho(s | \vx)$ that satisfies $A^{\vx}\rho = 0$.

For the coupled process, where both $s$ and $\vx$ are random, the evolution of the joint probability density function $P(s, \vx, t)$ is governed by the Master equation,
\begin{equation}
  \label{eq:4}
  \pd{}{t}P(s, \vx, t | s_{0}, \vx_{0}, t_{0}) = {\cal L}_{\epsilon}P \equiv \frac{1}{\epsilon}A^{\vx}P + L_{\vx}^{\epsilon}P.  
\end{equation}
In the limit $\epsilon \to 0$ we obtain the deterministic system,
\begin{equation}
  \label{eq:det}
  \frac{dx_{k}}{dt} = f_{k}(\vx) \equiv \sum_{s = 0}^{N_{s}}\rho(s | \vx)\sigma_{k}(s) - \gamma_{k} x_{k}.
\end{equation}
Fixed points are written as $\vx_{\rm fp}$, defined by $f_{k}(\vx_{\rm fp}) = 0$, for all $k=1,\cdots, d$.

\section{WKB analysis}
\label{sec:wkb}
First we consider a special case, one that is well known in the chemical literature \cite{doi76a,peliti85a}.
Suppose the gene state is fixed constant.
The WKB approximation of the stationary probability density function is obtained using the anzatz $\rmp(\vx) \lsim e^{-\Phi(\vx)/\epsilon}$.
Substituting the anzatz into \eqref{eq:3} and collecting leading order terms in $\epsilon$ yields the PDE $H(s, \vx, \vp) = 0$, where $\vp \equiv \nabla_{\vx}\Phi$ and
\begin{equation}
  \label{eq:12}
    H(s, \vx, \vp) \equiv \sum_{k = 1}^{d}\left[\sigma_{k}(s)(e^{p_{k}} - 1) + \gamma_{k} x_{k} (e^{-p_{k}}-1)\right].
\end{equation}

For the coupled process, where $s$ is also random, the situation is more complicated.
A popular approach (which is known to cause significant errors \cite{newby13a,zhang14a}) is to take the adiabatic limit and use the averaged Hamiltonian $\bar{H}(\vx, \vp) = H(\sum_{s}s\rho(s|\vx), \vx, \vp)$.

A much more accurate approximation can be obtained as follows.
The WKB approximation of the stationary probability density function is obtained using the anzatz
\begin{equation}
  \label{eq:wkb}
  \rmp(s, \vx) = \mathcal{N}K(\vx)r(s | \vx)e^{-\Phi(\vx)/\epsilon},
\end{equation}
where $\mathcal{N}$ is a normalization factor, $r(s | \vx)$ is the conditional distribution of the gene state given $\vx$, and $\Phi(\vx)$ is called the {\em quasipotential}.
The pre exponential factor $K(\vx)$ can be interpreted as a normalization factor for $r$ and is determined at higher order.
For completeness, we provide the details on how to compute $K(\vx)$ in Appendix \ref{sec:pef}.

Substituting \eqref{eq:wkb} into $\mathcal{L}^{\epsilon}_{\vx}P = 0$ and collecting leading order terms in $\epsilon$ yields
\begin{equation}
  \label{eq:1}
  \sum_{s' = 0}^{N_{s}}[A_{s s'}^{\vx} +  \delta_{ss'}H(s, \vx, \Phi'(\vx))]r(s' | \vx) = 0.
\end{equation}
This can be rewritten in matrix form as
\begin{equation}
  \label{eq:26}
  Qr = [A^{\vx} + \diag{H(s, \vx, \nabla_{\vx}\Phi(\vx))}]r = 0.
\end{equation}
It is not immediately clear what the Hamiltonian is since the above expression depends on $r$, which is also unknown.
One idea is to define the Hamiltonian to be $\mathcal{H}(\vx, \vp) = \det(Q)$.
But this definition is not the only one that we can make, for example, we could also define the Hamiltonian to be an eigenvalue of $Q$.
Then $r$ is simply the eigenvector (which we hope is positive since it approximates a probability distribution) corresponding to a zero eigenvalue.
Both of these definitions should yield the same WKB approximation if we can solve the static Hamilton-Jacobi equation, $\mathcal{H}(\vx, \nabla_{\vx}\Phi) = 0$.
The solution to the Hamilton-Jacobi equation can be written in terms of the method of characteristics \cite{ockendon03a}.
Setting $\vp = \nabla_{\vx}\Phi$, the Hamilton-Jacobi equation can be rewritten as the system of ODEs,
\begin{equation}
  \label{eq:rays}
  \dot{\vx} = \nabla_{\vp}\mathcal{H}(\vx, \vp),\quad \dot{\vp} = -\nabla_{\vx}\mathcal{H}(\vx, \vp).
\end{equation}
Now we ask a simple question.
Are MPPs also solutions to \eqref{eq:rays}?
The answer is certainty not evident from the WKB expansion.

We are then faced with a new question.
If we have multiple definitions of $\mathcal{H}$ that yield the same WKB approximation, do they also yield the same characteristics, and if the characteristics are different, which set correspond to MPPs?
Clearly, the solution surface $\nabla_{\vx}\Phi(\vx(t, \theta))$ can have different parameterizations even with the same initial data (i.e., the curves of constant $\theta$ need not be the same for different choices of $\mathcal{H}$).
This ambiguity is resolved in Section \ref{sec:ldp} using a large deviation principle, and the correct choice is to define ${\cal H}$ as the principal eigenvalue of $Q$;  it is real and simple, it is larger than the real part of the remaining eigenvalues, and its eigenvector $r$ is strictly positive.
The {\em Perron--Frobenius Theorem} guarantees that the principal eigenvalue exists with $r>0$ for all $p$ \cite{newby14mlpp}.
Moreover, it follows from the WKB approximation \eqref{eq:wkb} that the conditional probability distribution for the gene state is just the appropriately normalized right eigenvector (evaluated at $\vp = \nabla\Phi(\vx)$).

The correct choice of $\mathcal{H}$ also has significant consequences for certain numerical algorithms for computing MPPs and the quasipotential $\Phi$ that are based on the large deviation principle \cite{heymann08a,vanden-eijnden08a,cameron12a,newby14mlpp}.
Another shortcoming of making the wrong choice for $\mathcal{H}$ is that these algorithms depend on $\mathcal{H}(\vx, \vp)$ being a convex function of $\vp$.
Choosing $\mathcal{H}$ according to the large deviation principle guarantees that it is convex.
One way to compute MPPs and $\Phi$ is to use a numerical ODE integrator to solve \eqref{eq:rays}, and this method does not require $\mathcal{H}$ be convex \cite{newby12a,lu14a}.
However, there are two significant drawbacks of this method.
First, there is little control over which points $\vx$ the solution is evaluated at; often one is interested in approximating $\Phi$ on a discrete grid.
Each MPP is generated with a different initial condition and a shooting method is required to find the MPP with a specific end point.
Second, the shooting method rapidly becomes computationally expensive as the dimension $d$ increases.
A better choice for computing MPPs is the {\em geometric minimum action method} (GMAM), which is an iterative scheme to compute the MPP between a given start and end point \cite{heymann08a,vanden-eijnden08a}.
The quasipotential $\Phi(\vx)$ can be computed on a discrete grid with the {\em ordered upwind method} (OUM), which is a finite difference scheme for solving the Hamilton-Jacobi equation \cite{cameron12a,newby14mlpp}.
The GMAM and OUM requires $\mathcal{H}(\vx, \vp)$ and its various first and second order partial derivatives.

Because it is an eigenvalue of a matrix, it is not possible in general to explicitly calculate $\mathcal{H}$ except is special cases. 
However, there is an efficient iterative scheme, called the Collatz method, to numerically compute the principal eigenvalue, the details are included in Appendix \ref{sec:collatz}.
If we can numerically compute $\mathcal{H}$ then we need only know the coefficients in the characteristic polynomial of $Q$ (which can be explicitly calculated in general) to evaluate the various partial derivatives of $\mathcal{H}$. 
Write the characteristic equation for $Q$ as $\Omega(\lambda) \equiv a_{N_{s}}\lambda^{N_{s}} + \cdots + a_{1}\lambda + a_{0} = 0$, where the coefficients $a_{j}$ are functions of $\vx, \vp$.
Differentiating the characteristic equation and setting $\lambda = \mathcal{H}$ yields
\begin{equation}
  \label{eq:27}
   \nabla\mathcal{H}(\vx, \vp) = -\frac{\sum_{n=0}^{N_{s}}\mathcal{H}^{n}\nabla a_{n}(\vx, \vp)}{\sum_{n=1}^{N_{s}}n\mathcal{H}^{n-1}a_{n}(\vx, \vp)},
\end{equation}
Higher derivatives can be similarly defined.
We have found in practice that simply setting $\mathcal{H}=0$ in the above formula seems to yield well behaved numerical solutions, but justifying this choice in terms of stability requires more investigation.

\section{Deriving the large deviation principle}
\label{sec:ldp}
Large deviation estimates are based on the idea of formulating Laplace's method specifically for probability theory.
Recall that Laplace's method is used to approximate certain integrals.
For example, suppose we have a positive function $F(x)$ for which there is a single minimum at $x_{m}$ such that $F'(x_{m}) = 0$ with $\lim_{x\to \pm \infty}F(x) = \infty$.
Laplace's method yields the approximation
\begin{equation*}
  \int_{-\infty}^{\infty}e^{-F(x)/\epsilon} dx \lsim e^{- F(x_{m})/\epsilon}.
\end{equation*}
The idea is that the integral takes its largest contribution at the point $x_{m}$ where $F(x)$ is at its minimum because the integrand is exponentially decreasing away from $x_{m}$.

To see how this works in the context of probability theory, consider the random variable $X_{\epsilon}$.
We want to approximate the distribution with
\begin{equation}
  \label{eq:7}
  {\rm Pr}[X_{\epsilon}\in(x, x+dx) ] \lsim dx\, e^{-L(x)/\epsilon},
\end{equation}
for some function $L(x)$.
Let $p \in \mathbb{R}$.
Using \eqref{eq:7} we notice that
\begin{equation*}
  \ave{\explr{\frac{p X_{\epsilon}}{\epsilon} }} \lsim \int_{-\infty}^{\infty}\explr{\frac{p x - L(x)}{\epsilon} }dx.
\end{equation*}
Using Laplace's method, we know that the above integral obtains its largest contribution at $h(p) = \sup_{x \in \mathbb{R}}\{ xp - L(x)\}$ so that $\ave{\exp\{p X_{\epsilon}/\epsilon \}} \lsim \exp\{ h(p)/\epsilon\}$.
Hence, we can define the function,
\begin{equation}
  \label{eq:42}
  h(p) \equiv \lim_{\epsilon \to 0}\epsilon \log \ave{\explr{\frac{p X_{\epsilon}}{\epsilon} }}.
\end{equation}
Assume that $h$ is continuous and convex for all $p\in \mathbb{R}$ with $h(0) = 0$.
Then, we can define the function $L$ and its relationship to $h$ through the {\em Legendre transform} $\mathcal{T}$,
\begin{align}
  \label{eq:19}
  L(x) &\equiv \mathcal{T}[h(p)] = \sup_{p\in\mathbb{R}}\{xp - h(p) \} \\
  \label{eq:20}
  h(p) &= \mathcal{T}[L(x)] = \sup_{x \in \mathbb{R}}\{ xp - L(x)\}.
\end{align}
If $h(p)$ is a convex function of $p$ then $L$ is a convex function of $x$, and their derivatives are monotonic functions.
After differentiating each of the above expressions, we observe that the maximizers for \eqref{eq:19} and \eqref{eq:20} are given implicitly by
$x = h'(p)$,  $p = L'(x)$, respectively.
Hence, the procedure for Laplace's method is to first compute $h$ using \eqref{eq:42}.
If the limit exists then the probability density function can be approximated by \eqref{eq:7}, where $L(x)$ is given by \eqref{eq:19}.
This procedure can be generalized to apply Laplace's method to path integrals.

For simplicity, assume that there is a single slow species so that $d=1$ and $0 \leq x < \infty$.
The generalization for multiple slow species is presented in Section \ref{sec:mult-slow-spec}.
Discretize time with $t_{j} = t_{0} + j \Dt$, $j = 0,\cdots, N+1$.  
Let $\{s_{j},\,x_{j}\}$ be a discretized path where $s_{j} = s(t_{j})$ and $x_{j} = x(t_{j})$.
Assume that the end points ($j=0$ and $j=N+1$) of the path are fixed with $s_{N+1} = s$ and $x_{N+1} = x$.

The probability density function satisfying the Master equation can be written in terms of a path integral, which is a slightly different version of the standard integral form of the Chapman--Kolmogorov equation \cite{gardiner83a}.
If the path consists of a single interior point (i.e., $N=1$) then we have the Chapman--Kolmogorov equation,
\begin{equation}
  \label{eq:124}
  \rmp(s, x, t | s_{0}, x_{0}, t_{0}) = \sum_{s_{1}=0}^{N_{s}}\int_{0}^{\infty}dx_{1}\,\rmp(s, x, t |s_{1}, x_{1}, t_{1}) \rmp(s_{1}, x_{1}, t_{1} | s_{0} x_{0}, t_{0}).
\end{equation}
Compounding over multiple interior points along a path ($N>1$), we have
\begin{equation}
  \label{eq:43}
  \rmp(s, x, t | s_{0}, x_{0}, t_{0})  = \sum_{s_{1}, \cdots, s_{N}}\int_{0}^{\infty}dx_{1}\cdots\int_{0}^{\infty}dx_{N}\, \mathcal{P}[ \{s_{j}\},\{x_{j}\}],
\end{equation}
where $\mathcal{P}$ is the joint distribution over the path.
Using the Markov property, the path distribution can be written as the product
\begin{equation}
  \label{eq:129}
  \mathcal{P}[ \{s_{j}\},\{x_{j}\}, \{t_{j}\}] = \prod_{j=1}^{N}\rmp(s_{j}, x_{j}, t_{j} | s_{j-1}, x_{j-1}, t_{j-1}).
\end{equation}
We are interested in ``averaging out'' (not to be confused with stochastic averaging or adiabatic reduction) the dependence on $s$, hence we seek $$\mathcal{P}[\{x_{j}\}] = \sum_{s_{0},\cdots, s_{N+1}} \mathcal{P}[ \{s_{j}\},\{x_{j}\}].$$

Formally we take the continuum limit $N\to\infty$ with $\Dt \to 0$, and we assume that the random gene state $s(T)$ evolves on the fast timescale $T = t/\epsilon$.
We further assume that $x(t)$ and $p(t)$ are slowly varying functions on the fast timescale (i.e., $x'(t)$ and $p'(t)$ are $O(1)$).
The marginal path distribution $\mathcal{P}[x(t)]$ can be approximated as follows.
The generalization of \eqref{eq:42} for the path distribution is
\begin{equation}
  \label{eq:23}
  \mathcal{H}(x, p) = \lim_{\epsilon\to 0}\epsilon \log \ave{\explr{\int_{0}^{t/\epsilon}H(s(\tau), x, p)d\tau}},
\end{equation}
where
\begin{equation*}
  H(s, x, p) = \sigma(s)(e^{p} - 1) + \gamma x (e^{-p}-1).
\end{equation*}
Note that \eqref{eq:23} is mathematically rigorous for the case where there is no noise in the slow process (i.e., the protein synthesis and degradation is deterministic) and for the adiabatic limit (i.e., no noise from gene switching).
As formally justified in Appendix \ref{sec:laplaces-method-path}, \eqref{eq:23} is a straightforward generalization for the case where both noise sources are present.

It is worth taking a moment to compare \eqref{eq:23} to the adiabatic result.
The adiabatic limit yields a very different result, namely $\bar{H}(x, p) = \lim_{\epsilon \to 0}H(\ave{s(t/\epsilon)}, x, p)$.
In general, using the adiabatic limit or a diffusion approximation yields a Hamiltonian that agrees with \eqref{eq:23} only for $\abs{p} \ll 1$.

The generalization of \eqref{eq:19} is
\begin{equation}
  \label{eq:11}
  L[x, \dot{x}] = \sup_{p\in\mathbb{R}}\left[\dot{x} p - \mathcal{H}(x, p)\right],\quad \mathcal{H}(x, p) = \sup_{\dot{x}\in\mathbb{R}}\left[\dot{x} p - L[x, \dot{x}]\right],
\end{equation}
where $L$ is called the {\em Lagrangian} and the variable $p$ is called the {\em conjugate momentum}.
The maximizers are defined implicitly by
\begin{equation}
  \label{eq:16}
  \dot{x} = \pd{}{p}\mathcal{H}(x, p), \quad p = \pd{}{\dot{x}}L[x, \dot{x}].
\end{equation}

Using \eqref{eq:23}, the generalization of \eqref{eq:7} for the marginal path distribution is the large deviation principle,
\begin{equation}
  \label{eq:10}
  \mathcal{P}[x(t)] \lsim \explr{-\frac{1}{\epsilon}\act[x(t)]} ,
\end{equation}
where $\act[x(t)] =  \int_{t_{0}}^{t}L[x(\tau), \dot{x} (\tau)]d\tau$ is called the {\em action functional}; it is a measure of how likely a given stochastic trajectory is.
It follows from the definition of $L$ and the property $\mathcal{H}(x, 0) = 0$ that $\frac{d}{dt}\act[x(t)] \geq 0$.
As the action increases, the likelihood of observing a stochastic trajectory decreases.

\subsection{Most probable paths and WKB}

Combining \eqref{eq:124} with \eqref{eq:10} we can approximate the marginal probability density function $p(x, t)$ with the path integral over the set of continuous functions in $\mathbf{C}_{x_{0}}^{x} = \{ x(t) \in \mathbf{C}_{1} : x(0) = x_{0}, x(t) = x \}$,
\begin{equation}
  \label{eq:14}
  \rmp(x, t | x_{0}, t_{0}) \lsim \int_{\mathbf{C}_{x_{0}}^{x}} \mathcal{D}[x(t)] \explr{-\frac{1}{\epsilon}\act[x(t)]}.
\end{equation}
Rather than try to evaluate an integral on a function space, let us use Laplace's method to approximate the integral.
The largest contribution to the above path integral comes from the MPP, defined as the path that minimizes the action functional,
\begin{equation}
  \label{eq:55}
  x_{\rm MP}(t) \equiv \arginf_{x(t) \in \mathbf{C}_{x_{0}}^{x}}\int_{t_{0}}^{t}L[x(\tau), \dot{x} (\tau)]d\tau.
\end{equation}
Stochastic trajectories are highly likely to follow a MPP.
The probability of a $O(1)$ deviation from a MPP is exponentially small.
Using Laplace's method again, we have the approximation
\begin{equation}
  \label{eq:63}
    \rmp(x, t | x_{0}, t_{0})  \lsim \explr{- \frac{1}{\epsilon}\act[x_{\rm MP}(t)]}.
\end{equation}

Using the calculus of variations, one can show that the MPP defined by \eqref{eq:55} is a solution to Hamilton's equations \eqref{eq:rays}, derived using the WKB method!
That is, an MPP is a solution to the Euler--Lagrange equation,
\begin{equation}
  \label{eq:8}
  \frac{d}{dt}\pd{}{\dot{x}}L[x, \dot{x}] = \pd{}{x}L[x, \dot{x}].
\end{equation}
The above can be equivalently written as
\begin{equation}
  \frac{d}{dt}\left(\dot{x} \pd{}{\dot{x}}L[x, \dot{x}] - L[x, \dot{x}]\right) = 0.
\end{equation}
Using \eqref{eq:11}, we have the first integral,
\begin{equation}
  \dot{x}p - L[x, \dot{x}] = \mathcal{H}(x, p) = \text{Const}.
\end{equation}
Hence, $\mathcal{H}$ is constant along a MPP.~ 
The first equation in \eqref{eq:rays}, namely $\dot{x} = \pd{\mathcal{H}}{p}$, simply follows from \eqref{eq:16}.
The second equation is derived as follows.
Using the second equation in \eqref{eq:16} and \eqref{eq:8}, we have 
\begin{equation}
  \dot{p}  = \frac{d}{dt}\pd{}{\dot{x}}L[x, \dot{x}] = \pd{}{x}L[x, \dot{x}].
\end{equation}
We therefore need to show that $\pd{}{\dot{x}}L[x, \dot{x}] = -\pd{}{x}\mathcal{H}(x, p)$ on MPPs.
Using \eqref{eq:11} and writing the maximizer as $p(x)$ (i.e., the implicit solution of $\pd{}{x}\mathcal{H}(x, p) = \dot{x}$) as a function of $x$ yields
\begin{equation}
  \pd{}{\dot{x}}L[x, \dot{x}]= \pd{p}{x}\dot{x} - \pd{p}{x}\pd{}{p}\mathcal{H}(x, p(x)) - \pd{}{x}\mathcal{H}(x, p(x)).
\end{equation}
Then, it follows from the first equation in \eqref{eq:16} that $\pd{}{\dot{x}}L[x, \dot{x}] = - \pd{}{x}\mathcal{H}(x, p(x))$.

The final connection two the WKB solution comes by setting $\mathcal{H}=0$.
As noted above $\mathcal{H}=\text{Const}$ along MPPs.
If we restrict trajectories to the set of MPPs for which ${\cal H} = 0$, the action \eqref{eq:10} becomes
\begin{equation}
  \label{eq:134}
  \act[x_{\rm MP}(t)] = \int_{t_{0}}^{t} p(\tau)\dot{x}_{\rm MP}(\tau)d\tau.
\end{equation}
The connection between the large deviation principal and the WKB approximation \eqref{eq:wkb} of the stationary probability density function is as follows.
A MPP that starts at the stable fixed point $\vxa$ requires $t_{0}\to -\infty$ as $x_{0}\to \vxa$, because $\vxa$ is an unstable fixed point of \eqref{eq:rays} (even though it is a stable fixed point of the deterministic system). 
In fact, the deterministic system is recovered from \eqref{eq:rays} by setting $p = 0$, which means that deterministic flows exist on the stable manifold of the higher dimensional system.
Assume that the process is time autonomous so that the stationary density can be written as $\rmp(x) = \lim_{t_{0}\to -\infty}\rmp(x, t | \vxa, t_{0})$.  
We can approximate the stationary density by taking the limit $t_{0}\to-\infty$ and $x_{0}\to\vxa$ in \eqref{eq:63} and \eqref{eq:134} (assuming $x(t)$ and $p(t)$ satisfy \eqref{eq:rays}) to get
\begin{equation}
\label{eq:37}
  \rmp(x) \propto  \explr{-\frac{1}{\epsilon}\int_{-\infty}^{t}p(\tau) \dot{x}(\tau) d\tau} 
\equiv e^{-\Phi(x)/\epsilon},
\end{equation}
which is consistent with the WKB approximation \eqref{eq:wkb}.
Notice that the conjugate momentum $p$ can also be interpreted as $p = \frac{d\Phi}{dx}$.
It follows from \eqref{eq:134} and \eqref{eq:37} that the action determines the quasipotential with $\Phi(x(t)) = \act[x_{\rm MP}(t)]$.
Recall that $\act[x(t)]$ is a non decreasing function of time.
If $\dot{x}_{\rm MP}(t) \neq 0$ for some fixed $t$, it is easy to show that $\frac{d}{dt}\act[x_{\rm MP}(t)] = 0$ (and consequently $p = 0$) if and only if $\dot{x} = f(x)$.
In other words, the action and the quasipotential are flat when $x_{\rm MP}(t)$ is tangent to a deterministic trajectory.
This is always true at fixed points $x_{\rm fp}$.

\subsection{Multiple slow species}
\label{sec:mult-slow-spec}
For multiple slow species $\vx  \in \mathbb{R}^{d}$, $d > 1$, the Hamiltonian $\mathcal{H}(\vx, \vp)$ is a function of the vector $\vp = (p_{1}, \cdots, p_{d})$.
The Hamiltonian it is defined as the principal eigenvalue of the matrix $Q_{s, s'} = A^{\vx}_{s, s'} + \delta_{s, s'}H(s, \vx, \vp)$ and $r(s | \vx) > 0$ is its eigenvector.
The Lagrangian \eqref{eq:11} becomes
\begin{equation}
  \label{eq:13}
  L[\vx, \dot{\vx}] = \sup_{\vp \in \mathbb{R}^{d}}\{ \vp \cdot \dot{\vx} - \mathcal{H}(\vx, \vp)\}.
\end{equation}
MPPs satisfy Hamilton's equations \eqref{eq:rays}.

\section{Example: Toggle Switch}
\label{sec:tog}
\begin{figure}[htbp]
  \centering
  \includegraphics[width=11cm]{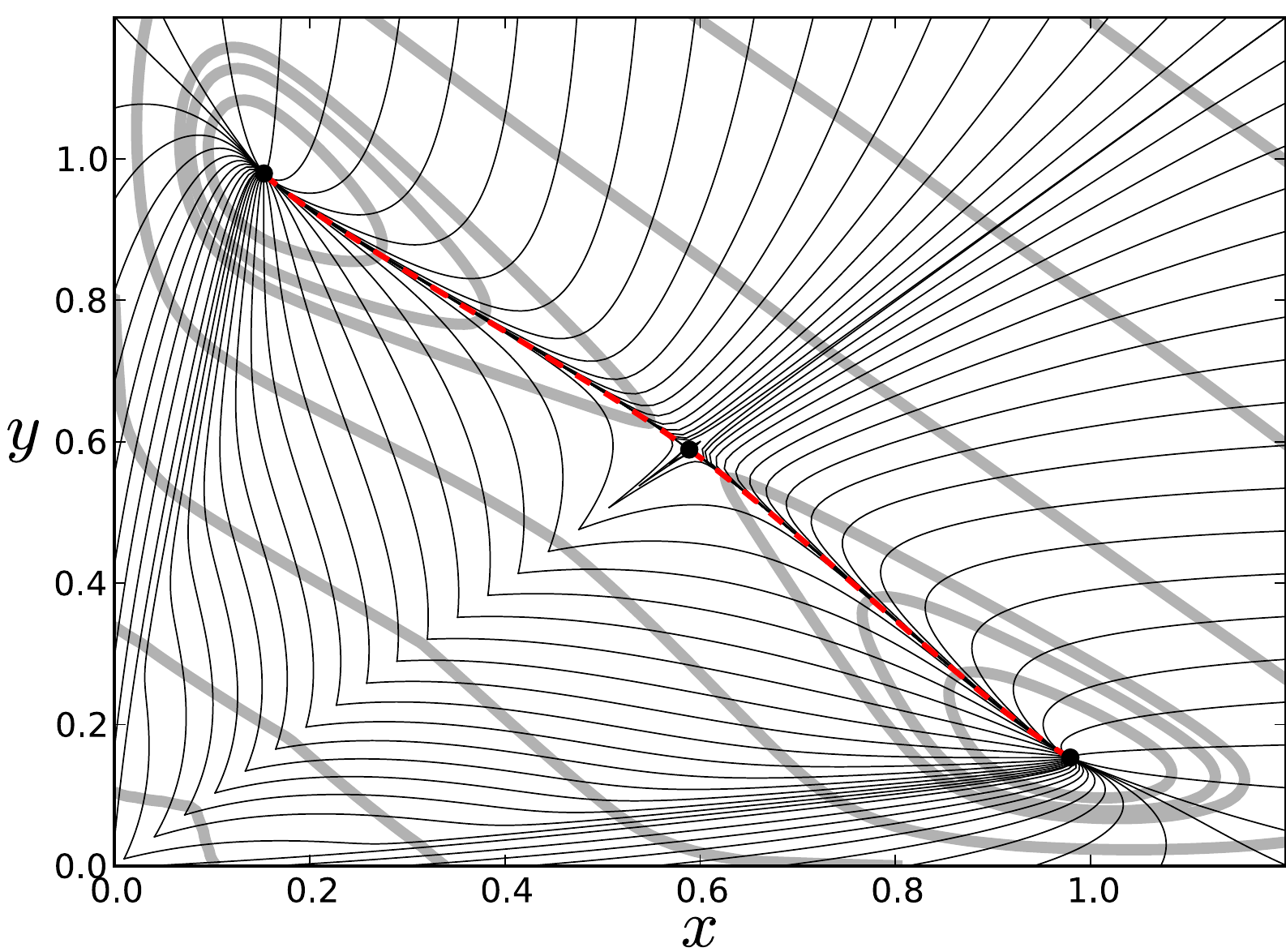}
  \caption{Several MPPs that start from each stable fixed point are shown (thin lines) for $\alpha = 2$, $b = 0.15$, and $\sigma = 1$. Red dashed lines show the MPEPs connecting the stable fixed points to the saddle. Level curves of the quasipotential $\Phi(x, y)$ are shown as thick lines.}
  \label{fig:rays}
\end{figure}

We turn now to a simple example to illustrate the results.
Consider two genes, $G_{X}$ and $G_{Y}$, responsible for producing proteins $X$ and $Y$, which have concentrations $x$ and $y$.
We assume that both genes can be turned on or one of the two can be turned off while the other is on.
Let $s = 1$ correspond to $G_{X}=\text{on}$, $G_{Y} = \text{off}$; let $s = 2$ correspond to $G_{X} = G_{Y} = \text{on}$; and let $s = 3$ correspond to $G_{X}=\text{off}$, $G_{Y} = \text{on}$.
Assume that a repressor formed by $X$ turns $G_{Y}$ off and a repressor formed by $Y$ turns $G_{X}$ off.
Once off, each gene turns back on at a rate $b$ when regulatory sequences unbind from the repressor.
The transition rate matrix for the gene state is,
\begin{equation}
    A^{(x,y)} =  \left[ \begin{array}{c c c}
   -b & x^{\alpha} & 0 \\
    b & -x^{\alpha}-y^{\alpha}& b \\
    0 & y^{\alpha} & -b
  \end{array}\right],
\end{equation}
where $\alpha$ is a parameter that controls the amount of nonlinearity in the dependence of the repressor concentration on protein concentration.
(For simplicity, we assume that the repressor concentrations are an instantaneous function of $x$ and $y$.)
The scaled protein synthesis rates are $\sigma_{X}(\{1, 2, 3\}) = \{0, \sigma, \sigma\}$ and $\sigma_{Y}(\{1, 2, 3\}) = \{\sigma, \sigma, 0\}$.
We set the characteristic time scale to the average lifetime of a protein so that the degradation rates are $\gamma_{X} = \gamma_{Y} = 1$.
For $\alpha\geq 2$ there is a range of values of $b$ for which the deterministic limit \eqref{eq:det} is bistable, having two stable fixed points $(x_{A}, y_{A})$ and $(y_{A}, x_{A})$, separated by an unstable saddle located on the center line $y = x$.
MPPs generated numerically using the GMAM are shown in Fig.~\ref{fig:rays} along with level curves of $\Phi(x, y)$ computed using the OUM.
The $x,y$ symmetry is evident from the level curves and MPPs.
Surrounding the two stable fixed points are closed level curves corresponding to a potential well.
The dashed curves show the MPP from the stable fixed points to the saddle located on the center line.
This curve is often referred to as the {\em most probable exit path} (MPEP) because it is the path most likely taken to transition from one of the stable fixed points to the other.
There are many other MPPs that reach the center line, but along the center line, the quasipotential is minimized at the saddle, which means that stochastic trajectories are most likely to cross the center line at the saddle.

\begin{figure}[htbp]
  \centering
  \includegraphics[width=12cm]{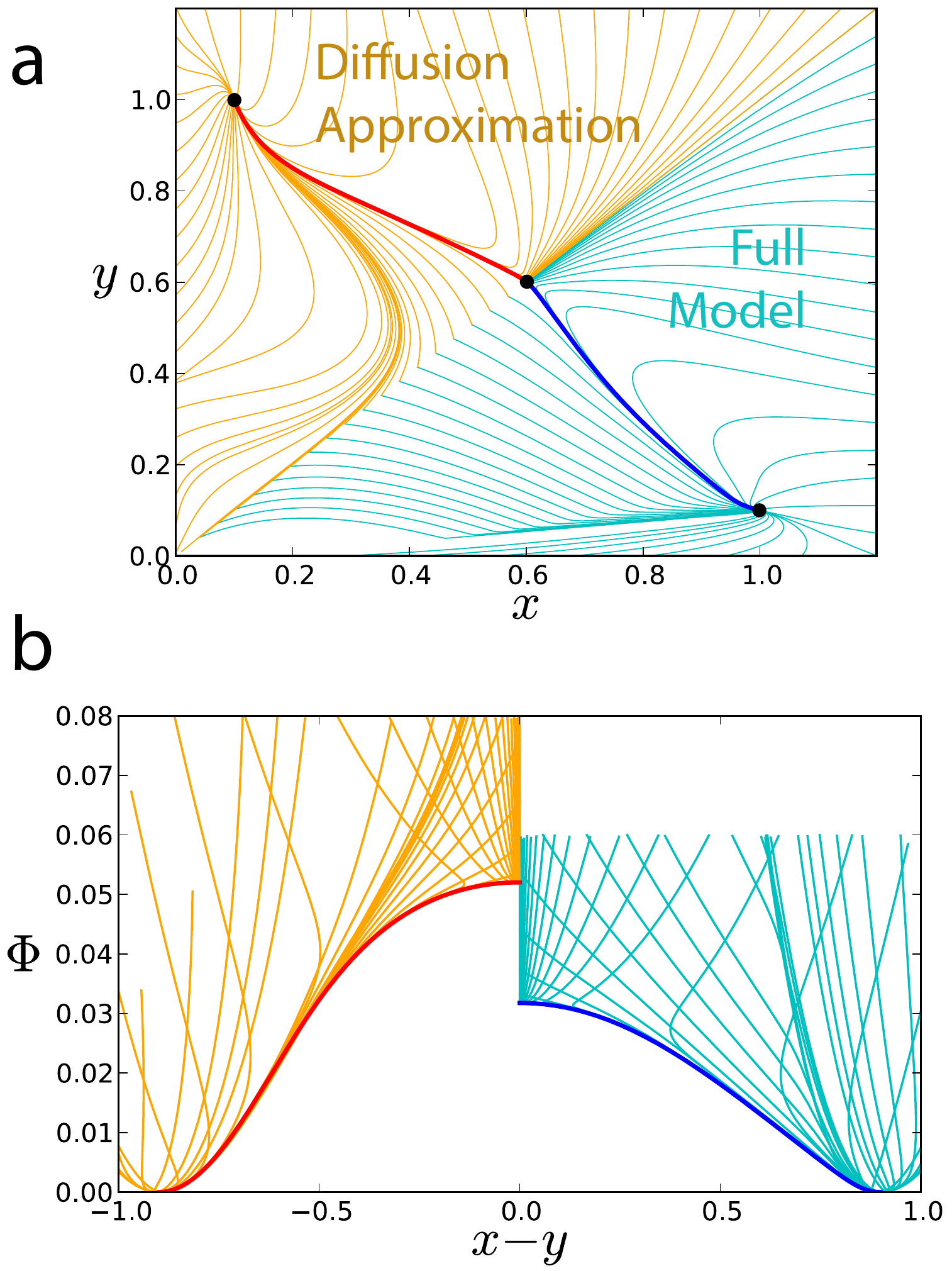}
  \caption{(a) MPPs computed using the GMAM for the full model ($y<x$) and for the diffusion approximation ($y>x$). The red and blue curves show the MPEP for the diffusion approximation and the full model, respectively. (b) The quasipotential along each of the MPPs shown in (a). Parameter values are $\alpha = 3$, $b = 0.15$, and $\sigma = 1$.}
  \label{fig:dacomp}
\end{figure}
It is interesting to see what differences a diffusion approximation has with these results.
A complete diffusion approximation of the toggle switch model, that includes noise effects from gene state switching and protein synthesis/degradation, was developed in \cite{kepler01a}.
In Fig.~\ref{fig:dacomp}(a) we show the MPPs derived from the full model in the lower half of the domain $y<x$ and the analogous MPPs obtained from the diffusion approximation on the upper $y>x$ portion of the domain.
We expect the diffusion approximation to be accurate in a small neighborhood of the stable fixed points.
After comparing the two approximations, we see that there are significant differences between the two sets of MPPs for small values of $x$ and $y$.
The MPEPs (red and blue curves in Fig.~\ref{fig:dacomp}) are also slightly different.

To understand the main source of error in the diffusion approximation, we examine the difference in the quasipotential.
Along a MPP, the quasipotential is given by
\begin{equation}
  \label{eq:21}
  \Phi(\vx(t)) = \int_{0}^{t}\vp(\tau)\cdot \dot{\vx}(\tau)d\tau,
\end{equation}
where $\vx(t)$ and $\vp(t)$ satisfy \eqref{eq:rays}.
In Fig.~\ref{fig:dacomp}(b) we show the quasi potential computed along each of the MPPs shown in Fig.~\ref{fig:dacomp}(a).
Notice a significant difference in the quasipotential at the saddle point, which has a exponentially large effect on the Kramers rate approximation of the mean exit time (not shown).

\section{Discussion}

We have derived a systematic, efficient, and practical extension of the QSA, which can be used to obtain reliable approximations of the stationary probability density function, MPPs, mean switching times, and the distribution of exit points.
We emphasize that our approach does not make the mistake of using an adiabatic limit or a diffusion approximation, which are unreliable and inaccurate for metastable events.
We have shown how the same asymptotic and numerical methods that exist to analyze a diffusion approximation can be applied to the full model.
We therefore conclude that {\em there is little to no benefit to be gained by using a diffusion approximation}.

The matrix $Q$ that determines $\mathcal{H}$ can be obtained using the simple and direct WKB approximation, and in situations where the principal eigenvalue of $Q$ cannot be explicitly calculated, an efficient numerical scheme can be used.
The correct formulation of $\mathcal{H}$ guarantees that it is a convex function of $\vp$, a property that is required for numerical schemes such as the OUM and GMAM.

\appendix
\section{Large deviation principal}
\label{sec:laplaces-method-path}
\newcommand{\tx}{X}
\newcommand{\ts}{S}
\renewcommand{\th}{\mathcal{H}}
For a model with only adiabatic-type weak noise, having no noise in the slow variable, \eqref{eq:23} becomes
\begin{equation}
  \label{eq:18}
    \mathcal{H}(x, p) = \lim_{\epsilon\to 0}\epsilon \log \ave{\explr{\int_{0}^{t/\epsilon}p f(S, x) d\tau}},
\end{equation}
where $\frac{dx}{dt} = f(s, x)$ is the deterministic motion for a given fixed state $s$.
The validity of the above is rigorously established in \cite{kifer09a} for the case where $S$ is a continuous Markov process, a Markov chain, or a combination of both.
For comparison, one can derive from the gene expression model a limiting process without noise in the slow variable.
By carefully rescaling the protein synthesis and degradation rates, one can derive a limiting process in which the protein concentration changes deterministically in between jumps in the gene state \cite{newby12a,newby13a}.
In this limit \eqref{eq:18} is recovered from \eqref{eq:23}.
That is, we have that $H(s, x, p) \to p f(s, x)$, which is consistent with the Hamiltonian of a deterministic process.

We state as a conjecture that \eqref{eq:23} is a straightforward generalization of \eqref{eq:18} for the case where there is noise in the slow variable.
In this section, we present a formal justification for this conjecture, using arguments that follow closely with the ideas found in the proof of \eqref{eq:18}.

We define the slow process as
\begin{equation*}
  X(t) = X(0) + \epsilon \left[Y^{+}(\frac{1}{\epsilon} \int_{0}^{t} \sigma(S(\epsilon^{-1}\tau))d\tau) - Y^{-}(\frac{\gamma}{\epsilon}\int_{0}^{t}X(\tau)d\tau)\right],
\end{equation*}
where $Y^{\pm}(t)$ are independent unit Poisson processes.
The fast process is $S(\epsilon^{-1}t)$, with transition rate matrix $A^{x}$.
Discretize the $\ts, x(t)$ dependent slow process with $\tx_{j} = \tx(t_{j})$, with increments $\Delta \tx_{j} = \tx_{j} - \tx_{j-1}$.
We seek an approximation of the marginal path distribution $\mathcal{P}[x(t)]$ having the form
\begin{equation}
  \label{eq:6}
  \mathcal{P}[\tx(t) \approx x(t)] \lsim \explr{-\frac{1}{\epsilon}\int_{0}^{t}L[x(\tau), \dot{x}(\tau)]d\tau}.
\end{equation}
Let $X'_{j} = \Delta X_{j}/\Delta t$.
Then using the discrete version of \eqref{eq:6} we have that
\begin{equation*}
  \ave{\explr{\frac{1}{\epsilon}\sum_{j = 1}^{N}p_{j} X'_{j}\Delta t}}  \lsim \int_{-\infty}^{\infty}dx'_{1}\cdots\int_{-\infty}^{\infty}dx'_{N}\explr{\frac{1}{\epsilon}\sum_{j=1}^{N}\left(p_{j}x'_{j} - L[x_{j}, x'_{j}]\right) \Dt}.
\end{equation*}
The largest contribution to each integral comes from
\begin{equation*}
 \th(x_{j}, p_{j}) = \sup_{x'_{j}\in \mathbb{R}}\left \{  p_{j}x'_{j} - L[x_{j}, x'_{j}] \right \},
\end{equation*}
Hence, the Hamiltonian should satisfy
\begin{equation}
  \label{eq:9}
  \int_{t_{0}}^{t}\mathcal{H}(x(\tau), p(\tau))d\tau = \lim_{\epsilon \to 0}\epsilon \log\ave{\explr{\frac{1}{\epsilon}\int_{t_{0}}^{t}p(\tau) X'(\tau)d\tau}}.
\end{equation}
A more useful expression than \eqref{eq:9} is derived as follows

First, we evaluate the mean over jumps in $\tx$ conditioned on $S$.
To leading order in $\Dt$, the propagator density function for $\Delta \tx$ (the probability density of jumping from $x$ to $x + y$ in the time interval $(t, t+\Dt)$) is
\begin{equation*}
  T(y | \Dt, s, x) \sim \delta(y) + \frac{\Dt}{\epsilon}\left[\sigma(s)(\delta(y - \epsilon) - \delta(y))  + \gamma x(\delta(y - \epsilon) - \delta(y)) \right].
\end{equation*}
Taking the mean of each $\exp\{\epsilon^{-1}p_{j}\Delta \tx_{j}\}$ over all $j = 0, \cdots, N$ yields
\begin{equation}
\label{eq:2}
\begin{split}
&\int_{-\infty}^{\infty}d y_{1}\cdots\int_{-\infty}^{\infty}d y_{N} \prod_{j = 1}^{N}T(y_{j} | \Dt, s, x_{j})e^{p_{j}y_{j}/\epsilon}  \\
&\quad= \prod_{j = 1}^{N}\left(1 + \frac{\Dt}{\epsilon}H(s, x_{j}, p_{j})\right) \sim \explr{\frac{1}{\epsilon}\sum_{j = 1}^{N}H(s, x_{j}, p_{j})\Dt} + o(\Dt),
\end{split}
\end{equation}
where
\begin{equation*}
  H(s, x, p) = \sigma(s)(e^{p} - 1) + \gamma x (e^{-p}-1).
\end{equation*}

To complete the evaluation of \eqref{eq:9}, we must average over the fast process $\ts(\epsilon^{-1}t)$.
Formally we take the continuum limit $N\to\infty$ with $\Dt \to 0$, and we assume that $x(t)$ and $p(t)$ are slowly varying functions on the fast timescale $T = t/\epsilon$.
In the continuum limit we have
\begin{equation}
  \label{eq:15}
  \lim_{\stackrel{N\to\infty}{\Dt \to 0}}\ave{\explr{\frac{1}{\epsilon}\sum_{j = 1}^{N}p_{j}\frac{\Delta \tx_{j}}{\Dt}\Dt}} \lsim \ave{\explr{\frac{1}{\epsilon}\int_{0}^{t}H(\ts(\epsilon^{-1}\tau), x(\tau), p(\tau))d\tau}}.
\end{equation}
Combining \eqref{eq:9} and \eqref{eq:15} yields,
\begin{equation}
  \label{eq:17}
    \int_{0}^{t}\th(x(\tau), p(\tau))d\tau = 
 \lim_{\epsilon\to 0}\epsilon \log \ave{\explr{\frac{1}{\epsilon}\int_{0}^{t}H(\ts(\epsilon^{-1}\tau), x(\tau), p(\tau))d\tau}}.
\end{equation}
Notice that \eqref{eq:17} differs drastically from the adiabatic result, which is essentially
\begin{equation*}
\begin{split}
  \int_{0}^{t}\bar{H}(x(\tau), p(\tau))d\tau &= \lim_{\epsilon\to 0}\epsilon \log \explr{\frac{1}{\epsilon}\int_{0}^{t}H(\ave{\ts(\epsilon^{-1}\tau)}, x(\tau), p(\tau))d\tau} \\
& = \int_{0}^{t}H(\lim_{T\to\infty}\ave{\ts(T)}, x(\tau), p(\tau))d\tau.
\end{split}
\end{equation*}
Then, differentiating both sides with respect to time yields
\begin{equation*}
  \bar{H}(x(\tau), p(\tau)) = H(\lim_{T\to\infty}\ave{\ts(T)}, x(\tau), p(\tau)),
\end{equation*}
where 
\begin{equation*}
  \lim_{T\to\infty}\ave{\ts(T)} = \sum_{s}s\rho(s | x), \quad \sum_{s'}A^{x}_{s, s'}\rho(s' | x) = 0,\quad\sum_{s}\rho(s | x) = 1.
\end{equation*}

Finally, we note that the formula \eqref{eq:23} given in the previous section is much simpler than \eqref{eq:17}.
A formal derivation of this formula using a multiple timescales argument is as follows.
Change variables to the fast timescale $T = t/\epsilon$ to get
\begin{equation*}
    \epsilon \int_{0}^{T}\th(x(t), p(t))dT' \sim  
 \epsilon \log \ave{\explr{\int_{0}^{T}H(\ts(T'), x(t), p(t))dT'}},
\end{equation*}
For fixed $t$ we have that $T\to\infty$ as $\epsilon\to 0$.
Divide through by $\epsilon T$, set $x(t) = x$ and $p(t) = p$, and take the limit $\epsilon\to 0$ with $t$ fixed to get
\begin{equation*}
  \th(x, p) = \lim_{T\to\infty}\frac{1}{T} \log \ave{\explr{\int_{0}^{T}H(\ts(T'), x, p)dT'}}.
\end{equation*}
The above mean assumes that $x$ is fixed constant so that
\begin{equation}
  \label{eq:1}
  \th(x, p) = \lim_{T\to\infty}\frac{1}{t} \log(\sum_{s'}U_{s, s'}(t)),
\end{equation}
where $U(t) = \exp\{tQ^{T}\}$ and $Q = A^{x} + \diag{H(s, x, p)}$.

We now show that \eqref{eq:1} converges to the principal eigenvalue of the matrix $Q$.
Although a general proof is only a few lines (see \cite{freidlin12a}), we offer a more descriptive argument, valid for the case where $Q$ is diagonalizable.
Suppose that the matrix $Q$ has eigenvalues $\lambda_{j}$, right eigenvectors $r_{j}$, and left eigenvectors $l_{j}$.
Let the principal eigenvalue be $\lambda_{0}$ with $r_{0}>0$, $l_{0} > 0$, and $\lambda_{0}>\Re(\lambda_{j})$, $j = 1,\cdots, N_{s}$.
Define the constants $c_{j} = \bfone^{T}r_{j}$.
Notice that $c_{0} > 0$ since $r_{0} > 0$.
We have
\begin{align*}
&\lim_{t\to \infty}\frac{1}{t}\log(\exp\{tQ^{T}\}\bfone) = \lim_{t\to \infty}\frac{1}{t}\log \sum_{j = 0}^{N_{s}}l_{j}e^{\lambda_{j}t}c_{j} \\
  & \quad= \lim_{t\to \infty}\frac{1}{t}\left[ \log  (\bfone e^{\lambda_{0}t}) + \log\left(l_{0}c_{0} + \sum_{j = 1}^{N}l_{j}e^{-(\lambda_{0} - \lambda_{j})t}c_{j}\right)\right] \\
  & \quad = \lambda_{0}\bfone  + \lim_{t\to \infty}\frac{1}{t}\log(l_{0}c_{0}) = \lambda_{0}\bfone.
\end{align*}

\section{The pre exponential factor}
\label{sec:pef}
From the leading order WKB solution we have $\Phi(\vx)$ and $r(s | \vx)$, which satisfies $\sum_{s'}Q_{ss'}r(s'|\vx) = 0$.
Define the left null vector $l(s | \vx)$ satisfying $\sum_{s'}Q_{s's}l(s'|\vx) = 0$, and assume that $l$ is normalized so that $\sum_{s}l(s | \vx)r(s | \vx) = 1$.
Collecting second order terms in the WKB expansion and applying a solvability condition yields the prefactor equation,
\begin{equation}
  \label{eq:150}
\begin{split}
  &\sum_{k = 1}^{d}\pd{K}{x_{k}}\sum_{s=1}^{N_{s}}l(s;\vx)\pd{J}{p_{k}}(s, \vx, \nabla \Phi) \\
&\quad = K\sum_{s=1}^{N_{s}}l(s|\vx)\sum_{k = 1}^{d}\left[\pcd{J}{p_{k}}{x_{k}}(s, \vx, \nabla \Phi)  + \frac{1}{2}\sum_{j}^{d}\pcd{\Phi}{x_{j}}{x_{k}}\pcd{J}{p_{j}}{p_{k}}(s, \vx, \nabla \Phi)\right]
\end{split}
\end{equation}
where
\begin{equation*}
  J(s, \vx, \vp) \equiv H(s, \vx, \vp)r(s | \vx).
\end{equation*}
Using the equation for the characteristics \eqref{eq:rays} we have that
\begin{equation}
  \frac{dx_{k}}{dt} = \pd{}{p_{k}}{\cal H}(\vx, \vp) =  \sum_{s=1}^{N_{s}}l(s;\vx)\pd{J}{p_{k}}(s, \vx, \nabla \Phi).
\end{equation}
Using the chain rule, we have
\begin{equation*}
  \frac{d}{dt}K(\vx(t)) = \sum_{k=1}^{d}\pd{K}{x_{k}}\frac{dx_{k}}{dt},
\end{equation*}
and it follows that along characteristics,
\begin{equation}
  \label{eq:56}
  \frac{dK}{dt} = K\sum_{s=1}^{N_{s}}l(s|\vx)\sum_{k = 1}^{d}\left[\pcd{J}{p_{k}}{x_{k}}(s, \vx, \vp)  + \frac{1}{2}\sum_{j}^{d}\pcd{\Phi}{x_{j}}{x_{k}}\pcd{J}{p_{j}}{p_{k}}(s, \vx, \vp)\right].
\end{equation}
Note that \eqref{eq:56} requires the Hessian matrix of $\Phi$, 
\begin{equation}
  \label{eq:3}
    Z_{j, k} \equiv \frac{\partial^{2}\Phi}{\partial x_{j} \partial x_{k}}.
\end{equation}
On characteristics, $Z$ satisfies the Ricatti equation \cite{ludwig75a,maier97a},
\begin{equation}
  \frac{dZ}{dt} = -ZDZ - ZC - C^{T}Z - G,
\end{equation}
where
\begin{equation}
  D_{j,k}(\vx, \vp) = \frac{\partial^{2}\mathcal{H}}{\partial p_{j}\partial p_{k}},\quad
  C_{j,k}(\vx, \vp) = \frac{\partial^{2}\mathcal{H}}{\partial p_{j}\partial x_{k}},\quad
  G_{j,k}(\vx, \vp) = \frac{\partial^{2}\mathcal{H}}{\partial x_{j}\partial x_{k}}.
\end{equation}

\section{Numerical algorithm to compute the principal eigenvalue}
\label{sec:collatz}
Power iteration is perhaps the simplest eigenvalue algorithm, and if it converges, it converges to the eigenvalue with the largest absolute value.
However, we seek the eigenvalue with the largest real part, which is not necessarily the eigenvalue with the largest absolute value.
Let $\kappa>0$ be a shift of the spectrum of $Q$, and define the matrix
\begin{equation}
  \label{eq:22}
  M_{\kappa} \equiv Q + \kappa I.
\end{equation}
For $\kappa$ sufficiently large, the matrix $M_{\kappa}$ is nonnegative and irreducible.
Both of these properties follow from the structure of the transition rate matrix $A^{x}$, which we assume to be irreducible.
It follows from the Perron--Frobenius Theorem that $M_{\kappa}$ has a simple real eigenvalue, call it $\mu$, that is greater in absolute value than the remaining eigenvalues.
The eigenvector corresponding to $\mu$, call it $r$, is strictly positive.
Setting $\mu = \kappa + \lambda$, we have that $M_{\kappa}r = Qr + \kappa r = \mu r = \kappa r + \lambda r$.
It follows that $\lambda$ and $r$ are an eigenpair of $Q$.
Moreover, $\lambda$ is the principal eigenvalue we seek.
We can then apply power iteration to $M_{\kappa}$, and the result converges to $\mu$ from which we obtain our desired result with $\lambda = \mu - \kappa$.
It remains to determine the convergence rate of the scheme and what value to chose for $\kappa$.

There are two relevant properties of nonnegative irreducible matrices.
Let $z$ be an arbitrary positive vector.
First, we have that $(M_{\kappa}z) >0$.
That is, the matrix maps positive vectors to positive vectors.
Second, it follows from Theorem 8.1.26 in Ref.~\cite{horn85a} that the eigenvalue $\mu$ is bounded by
\begin{equation}
  \label{eq:24}
  \min_{i} \frac{(M_{\kappa}z)_{i}}{z_{i}}\leq \mu \leq \max_{i}\frac{(M_{\kappa}z)_{i}}{z_{i}}.
\end{equation}
Moreover, if we optimize the bound over all possible positive vectors, we have
\begin{equation}
  \label{eq:25}
  \max_{z>0}\min_{i} \frac{(M_{\kappa}z)_{i}}{z_{i}} = \min_{z>0}\max_{i}\frac{(M_{\kappa}z)_{i}}{z_{i}} = \frac{(M_{\kappa}r)_{i}}{r_{i}} = \frac{\mu r_{i}}{r_{i}} = \mu.
\end{equation}
Let $\hat{\mu}$ be the next largest eigenvalue of $M$ in absolute value, and let $\hat{\mu} = \kappa + \hat{\lambda}$.
The asymptotic convergence rate is \cite{trefethen97a} 
\begin{equation*}
  \abs{\lambda - \lambda_{n}} = \abs{\mu - \mu_{n}} \sim O\left(\abs{\frac{\mu}{\hat{\mu}}}^{2n}\right) = O\left(\abs{\frac{\lambda + \kappa}{\hat{\lambda} + \kappa}}^{2n}\right),\quad n \gg 1.
\end{equation*}
It follows that as $\kappa \to \infty$, the convergence rate approaches unity.
We therefore want to choose the smallest possible value for $\kappa$, which corresponds to $\kappa = -\min_{i} Q_{ii}$.
Notice that the shift could be negative if all of the diagonal entries of $Q$ are positive, which corresponds to the case where $Q$ is already a nonegative matrix.
No shift is therefore necessary, but the negative shift speeds up convergence since $M_{\kappa}$ is also nonegative.
The algorithm written in Python code (using Scipy) as follows.
\begin{lstlisting}[frame=single]
from pylab import *
def collatz(Q, err_tol):
    ## Input: square matrix Q with dimension N 
    ##        error tolerance, err_tol
    ## Output: principal eigenvalue approximation, mu 
    N = M.shape[0]
    M = Q + kappa*eye(N)
    z = ones((N, 1))
    error = err_tol + 1.
    while (error > err_tol):
        q = dot(M, z)
        d = q/z
        a = min(d)
        b = max(d)
        mu = 0.5*(a + b)
        z = q/mu
        error = b - a
    return mu - kappa
\end{lstlisting}


\begin{thebibliography}{10}

\bibitem{assaf11a}
{\sc M.~Assaf, E.~Roberts, and Z.~Luthey-Schulten}, {\em Determining the
  stability of genetic switches: Explicitly accounting for mrna noise}, Phys.
  Rev. Lett., 106 (2011), p.~248102.

\bibitem{aurell02a}
{\sc E.~Aurell and K.~Sneppen}, {\em Epigenetics as a first exit problem},
  Phys. Rev. Lett., 88 (2002), p.~048101.

\bibitem{bressloff14a}
{\sc P.~C. Bressloff and J.~M. Newby}, {\em Path integrals and large deviations
  in stochastic hybrid systems}, Phys. Rev. E, 89 (2014), p.~042701.

\bibitem{cameron12a}
{\sc M.~Cameron}, {\em Finding the quasipotential for nongradient sdes},
  Physica D, 241 (2012), pp.~1532 -- 1550.

\bibitem{doering07a}
{\sc C.~R. Doering, K.~V. Sargsyan, L.~M. Sander, and E.~Vanden-Eijnden}, {\em
  Asymptotics of rare events in birth--death processes bypassing the exact
  solutions}, Journal of Physics: Condensed Matter, 19 (2007), p.~065145.

\bibitem{doi76a}
{\sc M.~Doi}, {\em Stochastic theory of diffusion-controlled reaction}, J.
  Phys. A: Math. Gen., 9 (1976), p.~1479.

\bibitem{eldar10a}
{\sc A.~Eldar and M.~B. Elowitz}, {\em Functional roles for noise in genetic
  circuits}, Nature, 467 (2010), pp.~167--173.

\bibitem{feng14a}
{\sc H.~D. Feng, K.~Zhang, and J.~Wang}, {\em Non-equilibrium transition atate
  rate theory}, J. Chem. Sci.,  (2014).

\bibitem{feng06a}
{\sc J.~Feng and T.~G. Kurtz}, {\em Large deviations for stochastic processes},
  vol.~v. 131 of Mathematical surveys and monographs, American Mathematical
  Society, Providence, R.I., 2006.

\bibitem{freidlin12a}
{\sc M.~I. Freidlin and A.~D. Wentzell}, {\em Random {P}erturbations of
  {D}ynamical {S}ystems}, Springer- Verlag, Berlin Heidelberg, 3rd edition~ed.,
  2012.

\bibitem{gardiner83a}
{\sc C.~W. Gardiner}, {\em Handbook of stochastic methods for physics,
  chemistry, and the natural sciences}, vol.~v. 13, Springer-Verlag, Berlin,
  1983.

\bibitem{heymann08a}
{\sc M.~Heymann and E.~Vanden-Eijnden}, {\em The geometric minimum action
  method: A least action principle on the space of curves}, Communications on
  Pure and Applied Mathematics, 61 (2008), pp.~1052--1117.

\bibitem{horn85a}
{\sc R.~A. Horn and C.~R. Johnson}, {\em Matrix analysis}, Cambridge University
  Press, Cambridge, 1985.

\bibitem{keener11a}
{\sc J.~P. Keener and J.~M. Newby}, {\em Perturbation analysis of spontaneous
  action potential initiation by stochastic ion channels}, Phys. Rev. E, 84
  (2011), p.~011918.

\bibitem{kepler01a}
{\sc T.~B. Kepler and T.~C. Elston}, {\em Stochasticity in transcriptional
  regulation: Origins, consequences, and mathematical representations},
  Biophys. J., 81 (2001), pp.~3116--3136.

\bibitem{kifer09a}
{\sc Y.~Kifer}, {\em Large deviations and adiabatic transitions for dynamical
  systems and markov processes in fully coupled averaging}, Memoirs of the
  American Mathematical Society, 201 (2009), pp.~1--144.

\bibitem{lu14a}
{\sc M.~Lu, J.~Onuchic, and E.~Ben-Jacob}, {\em Construction of an effective
  landscape for multistate genetic switches}, Phys. Rev. Lett., 113 (2014),
  p.~078102.

\bibitem{ludwig75a}
{\sc D.~Ludwig}, {\em Persistence of dynamical systems under random
  perturbations}, SIAM Review, 17 (1975), pp.~pp. 605--640.

\bibitem{maheshri07a}
{\sc N.~Maheshri and E.~K. O'Shea}, {\em Living with noisy genes: How cells
  function reliably with inherent variability in gene expression}, Annu. Rev.
  Biophys. Biomol. Struct., 36 (2007), pp.~413--434.

\bibitem{maier97a}
{\sc R.~S. Maier and D.~L. Stein}, {\em Limiting exit location distributions in
  the stochastic exit problem}, SIAM J. Appl. Math., 57 (1997), pp.~752--790.

\bibitem{newby14mrnapp}
{\sc J.~Newby}, {\em Bistable switching asymptotics for the self regulating
  gene}, ArXiv e-prints, 1407.4344 (2014), pp.~1--9.

\bibitem{newby14mlpp}
\leavevmode\vrule height 2pt depth -1.6pt width 23pt, {\em Spontaneous
  excitability in the morris--lecar model with ion channel noise}, ArXiv
  e-prints, 1406.2914 (2014), pp.~1--47.

\bibitem{newby13a}
{\sc J.~Newby and J.~Chapman}, {\em Metastable behavior in markov processes
  with internal states}, J. Math. Biol.,  (2013), pp.~1--36.

\bibitem{newby12a}
{\sc J.~M. Newby}, {\em Isolating intrinsic noise sources in a stochastic
  genetic switch}, Physical Biology, 9 (2012), p.~026002.

\bibitem{newby11b}
{\sc J.~M. Newby and J.~P. Keener}, {\em An asymptotic analysis of the
  spatially inhomogeneous velocity-jump process}, Multiscale Model. Simul., 9
  (2011), pp.~735--765.

\bibitem{ockendon03a}
{\sc J.~Ockendon, S.~Howison, A.~Lacey, and A.~Movchan}, {\em Applied partial
  differential equations}, Oxford University Press, Oxford, rev. ed~ed., 2003.

\bibitem{peliti85a}
{\sc L.~Peliti}, {\em Path integral approach to birth-death processes on a
  lattice}, J. Phys. France, 46 (1985), pp.~1469--1483.

\bibitem{sasai03a}
{\sc M.~Sasai and P.~G. Wolynes}, {\em Stochastic gene expression as a
  many-body problem}, Proc. Natl. Acad. Sci. U.S.A., 100 (2003),
  pp.~2374--2379.

\bibitem{schuss10a}
{\sc Z.~Schuss}, {\em Theory and applications of stochastic processes: an
  analytical approach}, vol.~v. 170 of Applied mathematical sciences, Springer,
  New York, 2010.

\bibitem{trefethen97a}
{\sc L.~N. Trefethen}, {\em Pseudospectra of linear operators}, SIAM Review, 39
  (1997), pp.~383--406.

\bibitem{vanden-eijnden08a}
{\sc E.~Vanden-Eijnden and M.~Heymann}, {\em The geometric minimum action
  method for computing minimum energy paths}, The Journal of Chemical Physics,
  128 (2008).

\bibitem{walczak05a}
{\sc A.~M. Walczak, J.~N. Onuchic, and P.~G. Wolynes}, {\em Absolute rate
  theories of epigenetic stability}, Proc. Natl. Acad. Sci. U.S.A., 102 (2005),
  pp.~18926--18931.

\bibitem{zhang14a}
{\sc B.~Zhang and P.~G. Wolynes}, {\em Stem cell differentiation as a many-body
  problem}, Proc. Natl. Acad. Sci. U.S.A., 111 (2014), pp.~10185--10190.

\end{thebibliography}
\end{document}